\documentclass[12pt]{iopart}
\expandafter\let\csname equation*\endcsname\relax
\expandafter\let\csname endequation*\endcsname\relax
\usepackage{amsmath} % need for subequations
\usepackage{slashed}
\usepackage{braket}
\usepackage{bm}% bold math
\usepackage{graphicx} % need for figures
\usepackage{verbatim} % useful for program listings
\usepackage{color} % use if color is used in text
\usepackage{subfigure} % use for side-by-side figures
\usepackage{hyperref} % use for hypertext links, including those to external documents and URLs
\usepackage{longtable}
\usepackage{dcolumn}
\usepackage{parskip} 

\usepackage{xcolor}
\usepackage[justification=centering]{caption}

\begin{document}

\title[]{Anomaly in decay of $^{8}$Be and $^{4}$He - can an observed light boson mediate low energy nucleon-nucleon interactions ? } 

\author{Martin Veselsk\'y$^1$, Vlasios Petousis$^1$, Jozef Leja$^2$}
\address{$^1$Czech Technical University in Prague - Institute of Experimental and Applied Physics}
\address{$^2$Faculty of Mechanical Engineering - Slovak University of Technology in Bratislava}
\ead{Martin.Veselsky@utef.cvut.cz} 
%\ead{vlasios.petousis@cvut.cz} 
%\ead{jozef.leja@stuba.sk} 
 
\begin{abstract}
We present a hypothesis that the reported anomaly 
in the folding angle distribution of electron-positron pairs, 
emitted in the decay of the excited levels of nucleus $^{8}$Be and $^{4}$He, 
can be related to the cluster structure of the decaying state. 
In particular, we suggest that the potentially reported boson with rest mass $m_{X}$=17 MeV 
can mediate the nucleon-nucleon interaction at the low-energy regime of QCD, 
in the weakly bound cluster state p+$^{7}$Li, and p+$^{3}$H. 
We explore a possible equations of state 
of symmetric nuclear matter, corresponding to the vector meson mass $m_{v}$=17 MeV, 
obtained using relativistic mean field theory of nuclear force using QHD-I.
We find several relations for boson masses, based on concepts of chiral symmetry breaking.
Both model approaches, thus point towards apparent restoration 
of chiral symmetry in nucleon-nucleon interaction at large distances, 
possibly via bounce into false instanton vacuum.    
A possible existence of a boson with rest mass $m_{X}$=17 MeV in the decay of high lying 
excited states of $^{8}$Be and $^{4}$He, could shed new light in one of the 
long lasting open questions in nuclear physics. 
\end{abstract}

% Uncomment for keywords
%\vspace{2pc}
\noindent{\it Keywords}: Beryllium Anomaly, Instantons, QCD, QHD-I, Chiral Symmetry \\
% Uncomment for Submitted to journal title message
%\submitto{\JPG}
% Uncomment if a separate title page is required
%\maketitle 
% For two-column output uncomment the next line and choose [10pt] rather than [12pt] in the \documentclass declaration
%\ioptwocol

\section{Introduction}
In 2016 an article of Krasznahorkay et al. appeared \cite{PRLAttila}, where an anomaly in the angular
correlation of electron-positron decay of the 1$^{+}$ excited level of $^{8}$Be nucleus at
18.15 MeV, specifically reported enhancement at the folding angles close to 140 degrees, 
is interpreted as a signature of a decay via emission of neutral boson with the
mass around $m_{X}$=17 MeV. 
Since initial article \cite{PRLAttila}, similar effect was reported by the same 
group also for the lower 1$^{+}$ excited state of the $^{8}$Be at 17.6 MeV \cite{Attila2} 
and more recently in the 0$^{-}$ excited state of the $^{4}$He at 21.01 MeV \cite{Attila4He}. 
The interpretation of the anomaly by introducing a new bosonic particle is rather far-reaching, especially 
in the domain of low energy nuclear spectroscopy, which immediately 
leads to question why such particle would not be reported by particle physicists
already before. Modern accelerators provide enough energy to produce it
copiously. Even if it would be a weakly interacting (sterile) particle, still
it should be detectable in the energy balance of the reported event. In
principle such particle is not excluded (see e.g. \cite{Fedor}) but possibly such
nuclear decay could still be explained within nuclear physics domain.  

First of all, one needs to understand the structure of the light nuclei like $^{8}$Be and $^{4}$He. 
The nucleus $^{8}$Be is rather 
unique case, practically unstable, which can be considered as consisting of
two $\alpha$-particles. In fact, the ground state is located above the sum of masses 
of two $\alpha$-particles and it is stabilized only by the Coulomb barrier 
(maximum 1.5 MeV for l=0). Basic mode of decay of the ground state and of excited states up to
17 MeV is the decay into two $\alpha$-particles. Ratio of energies of first two
excited states suggests that it is a nuclear rotator. However, starting from the
levels above 17 MeV basic mode of decay becomes emission of protons, while 
emission of gamma-particles and electron-positron pairs is also reported,
and decay into two $\alpha$-particles is suppressed. 

This sounds quite strange,
since these states are well above the Coulomb barrier and there should be no
obstacle for $\alpha$-decay, unless something happens to structure of these
excited states. What could happen there ? The answer can be possibly derived from the fact, that the proton 
separation energy of $^{8}$Be is 17.25 MeV and the 1$^{+}$ excited states at 17.6 and 18.15 MeV can be cluster 
configurations such as proton and $^{7}$Li, again stabilized only by Coulomb barrier. 
Thus the excited states above 17 MeV are possibly molecular states, stabilized mainly by Coulomb barrier of a given cluster configuration. 
Recent reported observation of similar effect in the excited state of the $^{4}$He \cite{Attila4He} appears to support the role of cluster structure 
in the decay of 1$^{+}$ excited states of $^{8}$Be at 17.6 and 18.15 MeV.  
In the case of 0$^{-}$ excited state of the $^{4}$He at 21.01 MeV, again above proton 
separation energy of around 20.5 MeV, the $^{3}$H+p cluster structure 
appears to play a role, thus further supporting 
the comparable role of cluster configurations in the case of  $^{8}$Be. 

\section{Phenomenological analysis}
As discussed above,  
namely the cluster structure can be the key for emergence  
of the effect suggesting a possible existence of 17 MeV boson. If this is the case, 
the strong interaction of the proton with remaining cluster may be the source 
of the reported boson. Compared to scattering experiments, the nuclear molecule 
like $^{7}$Li or $^{3}$H+p can exist much longer by several orders of magnitude, up to the 
lifetime of the excited state. Under such conditions low energy degrees 
of freedom of QCD can be possibly manifested. Thus the reported signal 
suggesting the 17 MeV boson can provide evidence for long sought for and 
now almost forgotten exchange boson mediating specifically strong interaction among 
nucleons. This explanation can provide alternative to presently 
preferred explanations in terms of dark matter or exotic particles. In particular,  
a possible existence of the effect in the decay of 1$^{+}$ excited states of $^{8}$Be 
lead the authors \cite{Attila4He} to possible interpretation in terms of a dark photon particle \cite{DarkPh}, a vector boson. 
Since until today a relevant theory of nucleon-nucleon 
interaction such as low energy QCD does not exist it is hard to judge on constraints 
concerning rest mass of such bosonic particle in nucleon-nucleon interaction. Below we discuss 
possible scenarios and resulting quantitative relations on both nucleonic and quark level. 

\subsection{Quantum hadrodynamics}
One of the models of nuclear forces used in nuclear theory 
is the QFT model of quantum 
hadrodynamics (QHD), which assumes in its simplest variant QHD-I an interaction of 
nucleons with massive scalar and vector bosons. This model does not provide 
constraint of the mass of vector boson, but the ratio of its coupling g$_{v}$ to its rest 
mass m$_{v}$ is constrained by saturation density and binding energy. 
Typically the mass of $\omega$-meson is chosen as m$_{v}$, which leads to 
large values of g$_{v}$ around 10-15. In this respect, the rest mass of $m_{v}$=17 MeV 
would lead to g$_{v}$ around or below unity. 
Concerning the range of interaction 
mediated by such vector boson, using the uncertainty principle and 
assuming massless particle, the range would be around 5 fm. This of 
course would be much less for particle with a rest mass at low energy and thus 
the typical range of strong force around 1-2 fm can be surely reproduced. 
The choice of vector particle reflects properties of QHD model, where contribution 
of pseudoscalar particle cancels out due to odd parity. However the 
obtained coupling of vector boson can be close also to the coupling of a pseudoscalar particle - obtained from the chiral symmetry breaking -  
as it is in the case of $\omega$ and $\pi$-mesons. 
In order to verify above mentioned possibility, the energy density of symmetric nuclear matter can be calculated using the formula \cite{QHDI} 
 \begin{center}
\begin{equation}
\centering 
\begin{split}
\epsilon = \frac{g_{v}^2}{2 m_{v}^2} \rho_{N}^2 + \frac{g_{s}^2}{2 m_{s}^2} (m_{N}-m_{N}^{*})^2 + \frac{\kappa}{6 g_{s}^3} (m_{N}-m_{N}^{*})^3 \\
+ \frac{\lambda}{24 g_{s}^4} (m_{N}-m_{N}^{*})^4 + \frac{\gamma}{(2 \pi)^3} \int_{0}^{k_{F}} d^{3}k \sqrt{k^2 + (m_{N}^{*})^{2}}  \label{EQHDI}
\end{split}
\end{equation}
\end{center}

where $g_{s}$,$m_{s}$ are coupling and rest mass of scalar boson, 
$g_{v}$,$m_{v}$ are coupling and rest mass of vector boson, 
$\kappa$,$\lambda$ are couplings of cubic and quartic 
self-interaction of scalar boson, $m_{N}$,$m_{N}^{*}$ are 
rest mass of nucleon and its effective mass, 
$\rho_{N}$ is the nucleonic density, $k_{F}$ is Fermi 
momentum of nucleons at zero temperature and 
$\gamma$ is the degeneracy (with value $\gamma$=4 for symmetric 
nuclear matter and $\gamma$=2 for neutron matter). 
In the simplest variant, without scalar boson self-interaction, 
the energy density is sensitive only to the ratios $g_{s}/m_{s}$ 
and $g_{v}/m_{v}$. Usually, the mass of vector boson is chosen 
as the mass of experimentally observed $\omega$-meson ($m_{\omega}$=783 MeV) 
and the mass of the scalar boson is a free parameter, with the typical 
values around 500 MeV. The corresponding couplings $g_{s}$ and $g_{v}$ 
are constrained using the experimental values of binding energy of 
symmetric nuclear matter ($E/A$=-16 MeV) and nuclear saturation 
density ($\rho_{0}$=0.16$ fm^{-3}$), which are deduced from the 
masses and radii of finite nuclei. The final constraints are \cite{QHDI}
\begin{equation}
\begin{split}
(\frac{m_{N}}{m_{s}})^2 g_{s}^2 = 357.4 \\
(\frac{m_{N}}{m_{v}})^2 g_{v}^2 = 273.8 \label{QHDIRatios}
\end{split}
\end{equation}

which guarantee proper values of both properties of nuclear matter. 
Typical values of couplings $g_{s}$ and $g_{v}$ with the choice 
of $m_{v}$=783 MeV are between 10 and 20, what obviously excludes perturbative 
treatment. The choice of vector boson mass $m_{v}$=17 MeV allows 
to describe symmetric nuclear matter using values of couplings 
around 0.3, what could allow perturbative treatment in principle. 
In order to test the choice $m_{v}$=17 MeV one can start with 
cubic scalar boson self-interaction. A wide range of equations of state, 
reproducing the binding energy and saturation density of 
symmetric nuclear matter, was obtained using the range of values of coupling $\kappa$ 
between -0.0032 and 0.0032 MeV. Outside of this interval 
properties of the global energy minimum of symmetric nuclear matter can not be described. 
In particular, the incompressibility of the softest equation 
of state obtained using $m_{v}$=17 MeV, $m_{s}$=32.22 MeV,
$g_{v}$=0.249, $g_{s}$=0.589, and $\kappa$=0.0032 MeV 
reaches down to incompressibility K$_{0}$=330MeV, which is however higher than 
the values, given by experimental constraints from 
nuclear reactions and astrophysical data, in particular 
from recent data of binary neutron star mergers. 

In order to obtain more realistic values of incompressibility, 
it is necessary to implement 
also quartic self-interaction of scalar meson. 
Using as a starting point the most promising values 
of $\kappa$ from scan with cubic self-interaction term only, a range of the 
values of $\lambda$ was examined and ultimately 
regions of parameters were identified, allowing to 
obtain equations of state with the values of incompressibility 
in the range K$_{0}$=240-260 MeV, which was constrained e.g. 
in our recent study on fusion hindrance in reactions, 
leading to production of superheavy elements \cite{MVCoMDSHE}. 
Similar values were extracted recently \cite{BinNSMerg} from analysis 
of the binary neutron star merger event GW170817. 
Further constraints were applied based on the values 
of binding energy \cite{EoSFOPI} and pressure \cite{EoSAlam} 
of symmetric nuclear matter at double and triple 
of saturation density. The resulting parameter 
sets for equations of state with incompressibilities 
K$_0$=245, 250, and 260 MeV, respectively, are shown 
in the Table \ref{tab1}. 

In this respect it is worthwhile to mention 
that the value of second derivation of single particle 
density, from which is the value of K$_0$ derived, 
must not necessarily represent the behavior at high 
densities. This is true only if the shape of equation 
of state can be fitted using polynomial function. 
Then the relation between global minimum (saturation 
density) and behavior at high densities can be established. 
Fortunately this is the case here and the obtained equations 
of state can be described up to triple of saturation density 
using the quartic polynomial and the values of K$_0$, 
extracted around the saturation density are consistent with 
the values obtained from the global fit of equation of state 
up to triple of saturation density. 
Concerning the mass of scalar boson, the value around 
m$_s$=25.5 MeV was obtained. Until now, no such particle 
was reported, however also for usual 
value of m$_s$=500-520 MeV, corresponding to the choice m$_v$=783 MeV, 
this question is not solved conclusively, so the scalar boson 
with mass m$_s$=25.5 MeV can be considered as an artefact of the model. 

Concerning bosonic particle with mass m$_v$=17 MeV, it can be considered as specific 
to low-energy nucleon-nucleon interaction with no obvious  
relation to dark matter, introduced in cosmological models, 
or to new type of interaction. This bosonic particle 
can represent low-energy degrees of freedom of QCD, where 
however no conclusive theory exists at present.  
In this respect, an intuitive candidate for mediator of 
interaction between nucleons would be a complex of 
gluons, called usually glueball. However, mass of such 
elementary particle is estimated around 1.5-1.7 GeV. 
Still, such estimates possibly include mechanism of confinement, 
which generates more than 95 \% of the nucleonic mass. 
A possible observation of particle with mass around 17 MeV, 
comparable to quark current masses,  
may suggests that mechanism of confinement is not 
applicable and that the nucleons exchange virtual particles 
consisting of unconfined gluons or quarks, generated from QCD vacuum. 
In any case, the corresponding value of $g_{v}$=0.245  
appears compatible with such explanation and thus  
may provide some encouragement concerning possibility 
to formulate corresponding theory. 

Of course, the analysis, performed here does not 
exclude the role of heavier (confined) bosonic particles as mediators of nucleon-nucleon interaction.  
Both lighter and heavier bosonic particles can contribute 
to low-energy nucleon-nucleon interaction, but the interaction 
via lighter virtual bosons may be pronounced especially when nucleons 
interact at large distance, what would be the case if nuclear 
molecule is formed, as can be expected for highly excited 
states of $^{4}$He and $^{8}$Be. 
In any case, possible existence of light bosonic particle 
opens possibility to understand better the long standing open question 
of nuclear physics. 

Realistic relativistic mean field calculations of finite nuclei, like ones based on point-like couplings \cite{FiNuPoCou} and meson exchange \cite{mesonExc} effective Lagrangians, are complex tasks which goes far beyond the scope of the present work. We tested the obtained equations of states on less computation-complex finite object such as neutron star. We observe \cite{NeutronStarPaper} that the obtained simple equation of state, allows reasonable description of mass-radius diagram, even allowing for the existence of a neutron star up to 2.5 Solar masses, for which a possible candidate was observed in recent gravitational wave event GW190814 \cite{GW190814}. 

Being aware of connection between neutron star radius and neutron skin of nuclei such as $^{208}Pb$, we would anticipate that introduction of a light boson acting at longer distance would allow to harmonize the relativistic mean field results for neutron skin with recent experimental observations of PREX-2 \cite{PREX-2}. 
\begin{table}[ht]
\caption{Constrained parameter sets for equations of state with incompressibilities K$_0$=245, 250, and 260 MeV.}
\vskip 0.2cm
{\centering
\begin{tabular}{llllllllr}
\hline
K$_0$&m$_{v}$&m$_{s}$&g$_{v}$&g$_{s}$&$\kappa$&$\lambda$ \\
\hline
MeV & MeV& MeV& & & MeV& \\
\hline
245&17.&25.58&0.2407&0.4666&0.0039&-0.001396 \\
250&17.&25.58&0.2417&0.4703&0.00398&-0.001316 \\
260&17.&25.58&0.2417&0.4684&0.00374&-0.001204 \\
\hline
\end{tabular}
\par}
\label{tab1}
\end{table}

\subsection{Instantons and chiral symmetry}
The analysis, based on the model of QHD, is motivated by 
a possible existence of the signal of bosonic particle in the decay 
of 1$^{+}$ excited states of $^{8}$Be at 17.6 and 18.15 MeV 
and thus can suggest that it is a vector particle. 
However, recent reported observation of analogous effect in the decay 
of 0$^{-}$ excited state of the $^{4}$He at 21.01 MeV 
allows explanation in terms of pseudoscalar particle, 
while the case of  $^{8}$Be can be possibly explained by emission 
of such particle with orbital angular momentum l=1. 
There exists theory of nuclear force based on exchange 
of pseudoscalar particle, a pion. Pion is a hadronic 
particle observed in particle physics, and it is also related 
to both spontaneous and explicit breaking of chiral symmetry, with its 
mass of $m_{\pi}$=135 MeV being a measure of the latter effect. 
In this respect, the reported observation of pseudoscalar bosonic 
particle with rest mass of $m_{X}$=17 MeV might suggest that 
for whatever reason, the chiral symmetry is restored 
in the case of low energy nucleon-nucleon interaction 
at large distances. 

This is quite surprising possibility, 
which could be possibly related to properties of the 
QCD vacuum. There exists a theory of gluonic solutions 
of QCD equations called instantons \cite{Inst}. Such QCD vacuum 
consists of granular configuration, where instantons 
are regions with gluonic field with radius of 1/3 fm, 
with typical distance around 1 fm. These gluonic formations 
can be understood as tunneling events where quarks can 
be hopping between instantons, in analogy with conducting 
electrons in metals. The model of random 
instanton liquid allows to describe phenomenologically 
the experimental values of gluonic and quark condensates, 
and leads to chiral symmetry breaking  
(for review see e.g. \cite{RevDiak}, \cite{RevSchaShur}).  
It can be also used to derive the properties of 
mesons, in particular of the pseudoscalar meson. 
The formula for mass of pion derived 
using the instanton model reads \cite{tHooft} 
\begin{equation}
m_{\pi}^2 = \frac{m}{f}
\label{PionMass}
\end{equation}

where $m$ can be related directly to the mass of fermionic particle (quark) \cite{tHooft2} 
and $f$ is the expectation 
value of scalar field in the minimum of the Mexican hat 
potential. In traditional theory of chiral symmetry 
such as linear $\sigma$-model (see e.g. review \cite{Koch}), 
$m$ is interpreted as a measure of explicit violation of 
chiral symmetry, and the value of $f$ is taken equal to  
pion decay constant $f_{\pi}$=93 MeV, which is also equal 
to vacuum expectation value (VEV) of scalar field. 
Compared to the experimental pion mass around 
$m_{\pi}$=135 MeV the mass of pseudoscalar particle $m_{X}$=17 MeV 
means that, as two opposite extremes, either the $m$ (and thus possibly 
the quark mass $m_{q}$) gets 63 times smaller 
or the value of $f$ is 63 times larger. 
In this respect, it is remarkable that the ratio 
of dynamical quark mass around $m_{q,dyn}$=310 MeV to current quark 
mass of around $m_{q,curr}$=5 MeV fits this value almost perfectly:  
\begin{equation}
\frac{m_{X}^2} {m_{q,curr}} \simeq \frac{m_{\pi}^2} {m_{q,dyn}}
\label{MassScaling}
\end{equation}

This appears to signal that as explicit breaking of chiral symmetry 
is restored from dynamical mass scale down to current mass scale, 
and also the properties of pseudo-scalar particle get 
closer to Goldstone boson. In theory of chiral symmetry the axial current 
of pion is usually assumed to counterbalance the axial current of nucleon, 
so that total axial current is conserved (see also \cite{Koch}). 
The relation (\ref{MassScaling}) then suggests that the same 
outcome is reached when considering sum of axial current of current quark 
and axial current of 17 MeV boson. The 17 MeV boson thus appears 
to restore the symmetry of QCD in the vacuum, where current 
quarks are considered as a degree of freedom.  
Based on this, it is possible to derive an analogue 
of the Goldberger-Treiman relation \cite{G-T}
\begin{equation}
g_{Xqq} = \frac{g_{A} m_{q,curr}}{f_{\pi}}
\label{GTXqq}
\end{equation}

where $g_{Xqq}$ is the coupling of $X$-boson to current quark, 
$g_{A}$ is renormalization factor of axial current and 
$f_{\pi}$ was adopted as proportionality factor also for 
axial current of $X$-boson. The value of coupling $g_{Xqq}$=0.07 is obtained, 
what suggests the coupling with nucleon $g_{XNN}$=0.21.
In this respect, it is apparent 
that also the results using relativistic mean field theory 
of nuclear force, presented above, point practically in the same 
direction, since the value of coupling corresponding 
to vector mesonic mass of 17 MeV is $g_v$=0.24, 
while the value of coupling $g_v$=10-20, corresponding 
to vector meson mass of 783 MeV is in good agreement 
with value $g_{\pi NN}$=13 for experimental coupling of pion to nucleon, 
confirmed also by original Goldberger-Treiman relation. 
Similar value of coupling for low-energy QCD is actually obtained also 
in the calculation using the model of spaghetti vacuum \cite{SpgVac}. 
When adopting the value of $f_{\pi}$ in the denominator 
of eq. (\ref{PionMass}), the value of $m$, a quark condensate, drops 
from the value of $m$=(120 MeV)$^{3}$ for $m_{\pi}$=135 MeV 
down to the value of $m$=(30 MeV)$^{3}$ for $m_{X}$=17 MeV. 
Since the value of $m$ scales with quark mass, 
the relation in eq. (\ref{MassScaling}) depends only 
on vacuum expectation value $f_{\pi}$, what explains the validity of this relation. 

This situation depicted in the Fig.1, which compares two trends of mesonic versus quark masses. First trend combines formulas (\ref{QHDIRatios}) and (\ref{GTXqq}) under assumption that the obtained coupling correspond to ${{g_{XNN}} = {3g_{Xqq}}}$. Second trend represents formula (\ref{MassScaling}). Both trends cross around the value of $m_{X}$=17 MeV and $m_{q}$=5 MeV. 

Possible observation of a boson with mass $m_{X}$=17 MeV and 
the relations (\ref{MassScaling}) and (\ref{GTXqq}) thus suggests 
existence of a new scale in QCD, with chiral symmetry 
practically restored. 
\begin{figure}[h]
\centering
\includegraphics[width=160mm, height=100mm]{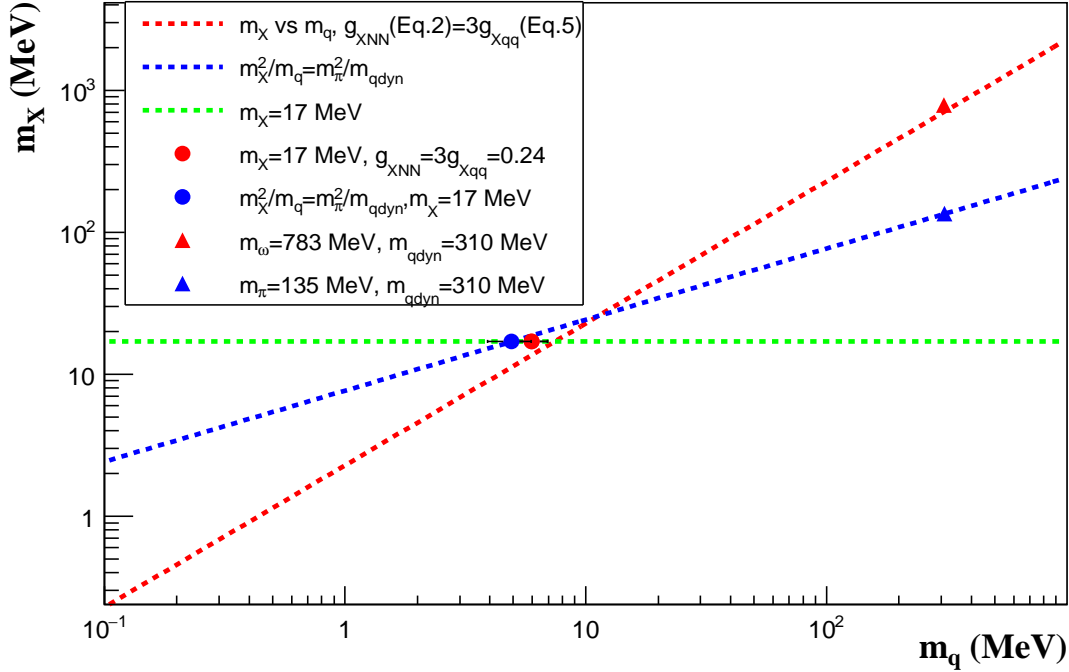}
\caption{(Color online). Trends of mesonic versus quark masses. First trend combines formulas (2) and (5) as explained in the text. Second trend represents formula (4).}
\end{figure}
The  reason for apparent restoration of chiral symmetry 
suggested above is a matter of discussion. It might be related to the 
interaction between nucleons at distance, mediated by instantons, 
where possibly the density of instantons mediating interaction 
in a region between nucleons becomes smaller and  
possibly a bounce into a false instanton vacuum (a quantum mechanical tunneling into a metastable vacuum and back), 
a mechanism based on instanton model \cite{Coleman}, can occur. 
Such a state might obviously differ from a stable QCD vacuum, 
which is satisfactorily described by the lattice QCD (for review see e.g. \cite{LQCD}). 
The relations (\ref{MassScaling}) and (\ref{GTXqq}) will 
characterize such state and  
the mechanism of explicit chiral 
symmetry breaking may be substantially weaker or nonexistent. 
Assumption of a bounce into a false vaccum state might sound as speculation, but 
one should remember that a long-lived practically 
unbound system of nucleons and clusters - a nuclear molecule -  
with sufficient excitation energy to produce particle 
with rest mass of 17 MeV, 
is a unique system, reported exclusively in highly 
excited states of light nuclei like $^{4}$He and $^{8}$Be. 
This might be practically the only opportunity to study long 
range interaction between nucleons on long timescale 
(compared to typical nuclear time of the order of 10$^{-23}$ s), 
which was not explored in detail yet and thus surprises 
can not be excluded.  

\section{Conclusions}
To summarize, we present a hypothesis that the anomaly 
in the folding angle distribution of electron-positron pairs, 
emitted in the decay of the excited level of nucleus $^{8}$Be at 18.15 MeV
\cite{PRLAttila}, can be related to the cluster structure of the decaying state. 
Furthermore, based on subsequent reported observation of similar anomaly 
in decay of excited state of $^{4}$He \cite{Attila4He}, which can 
be represented as a cluster state p+$^{3}$H, we present 
a hypothesis that the potentially reported boson with rest mass $m_{X}$=17 MeV  
can be an exchange boson mediating the nucleon-nucleon 
interaction at the low-energy regime of QCD. 
We also present a range of possible equations of state 
of symmetric nuclear matter based on this hypothesis 
and obtained using relativistic mean field theory of nuclear force, 
QHD-I in particular, including 
the equations of state of physical interest with incompressibilities K$_{0}$=240-260 MeV. 
The value of coupling $g_{v}$, corresponding to the meson masses $m_{v}$=17 MeV 
is lower than unity. Based on concepts of instanton liquid model 
with resulting chiral symmetry breaking, 
we show that reduction of the rest mass of pseudoscalar particle from physical 
value $m_{\pi}$=135~MeV to $m_{X}$=17 MeV is equivalent to reduction of the quark mass 
from dynamical value around 310 MeV down to current quark mass  
($\frac{m_{X}^2} {m_{q,curr}} \simeq \frac{m_{\pi}^2} {m_{q,dyn}}$). 
Assumption of conservation of axial current and resulting 
variant of Goldberger-Treiman relation leads to 
the value of coupling close to the 
results from relativistic mean field theory of nuclear force.  
Both model approaches thus point towards apparent restoration 
of chiral symmetry in nucleon-nucleon interaction at large distances, 
possibly via bounce into state of a false QCD vacuum, 
a mechanism based on the instanton model.    
Possible existence of boson with rest mass $m_{X}$=17 MeV in the decay of high lying 
excited states of $^{8}$Be and $^{4}$He, could shed new light in one of the 
long lasting open questions in nuclear physics. 

\section*{Ackwnoledgments}
This work is supported by the European Regional Development Fund-Project
Engineering applications of microworld physics, (Contr.No.CZ.02.1.01/0.0/0.0/16\_019/0000766).

\end{document}